\documentclass[preprint, aps, prb]{revtex4-1}
\usepackage{amsmath}    
\usepackage[dvipdfmx]{graphicx}   
\usepackage{verbatim}   
\usepackage{color}      
\usepackage{subfigure}  
\usepackage{hyperref}   
\usepackage{textcomp}
\usepackage{upgreek}

\usepackage{natbib}

\newcommand{\niso}{Ni$_2$InSbO$_6$}
\newcommand{\nto}{Ni$_3$TeO$_6$}
\newcommand{\qvec}{\mbox{\boldmath $q$}}
\newcommand{\Hvec}{\mbox{\boldmath $H$}}
\newcommand{\cvec}{\mbox{\boldmath $c$}}
\newcommand{\Hperpc}{\mbox{\boldmath $H$} $\perp$ \mbox{\boldmath $c$}}
\newcommand{\Hparac}{\mbox{\boldmath $H$} $\parallel$ \mbox{\boldmath $c$}}
\newcommand{\Hparaa}{\mbox{\boldmath $H$} $\parallel$ \mbox{\boldmath $a$}$^*$}
\newcommand{\Hparaq}{\mbox{\boldmath $H$} $\parallel$ \mbox{\boldmath $q$}}
\newcommand{\Hperpq}{\mbox{\boldmath $H$} $\perp$ \mbox{\boldmath $q$}}

\begin{document}

\title{Metamagnetic transitions and magnetoelectric responses in a chiral polar helimagnet \niso }
\author{Y.\,Araki,$^1$ T.\,Sato,$^1$ Y.\,Fujima,$^1$ N.\,Abe,$^1$ M.\,Tokunaga,$^2$ S.\,Kimura,$^3$ D.\,Morikawa,$^4$ V.\,Ukleev,$^4$ Y.\,Yamasaki,$^{4,5}$ C.\,Tabata,$^6$ H.\,Nakao,$^7$ Y.\,Murakami,$^7$ H.\,Sagayama,$^7$ K.\,Ohishi,$^8$ Y.\,Tokunaga,$^1$ and T.\,Arima$^{1,4}$}
\affiliation{$^1$Department of Advanced Materials Science, University of Tokyo, Kashiwa 277-8561, Japan}
\affiliation{$^2$Institute for Solid State Physics, University of Tokyo, Kashiwa 277-8581, Japan}
\affiliation{$^3$Institute for Materials Research, Tohoku University, Sendai 980-8577, Japan}
\affiliation{$^4$RIKEN Center for Emergent Matter Science (CEMS), Wako 351-0198, Japan}
\affiliation{$^5$Research and Service Division of Materials Data and Integrated System (MaDIS), National Institute for Materials Science (NIMS), Tsukuba 305-0047, Japan}
\affiliation{$^6$Institute of Materials Structure Science, High Energy Accelerator Research Organization, Tsukuba 305-0801, Japan}
\affiliation{$^7$Condensed Matter Research Center and Photon Factory, Institute of Materials Structure Science High Energy Accelerator Research Organization, Tsukuba 305-0801, Japan}
\affiliation{$^8$Neutron and Technology Center, Comprehensive Research Organization for Science and Society (CROSS),Tokai 319-1106, Japan}

\date{\today}

\begin{abstract}
Magnetic-field effect on the magnetic and electric properties in a chiral polar ordered corundum \niso\ has been investigated. Single-crystal soft x-ray and neutron diffraction measurements confirm long-wavelength magnetic modulation. The modulation direction tends to align along the magnetic field applied perpendicular to the polar axis, suggesting that the nearly proper-screw type helicoid should be formed below 77\,K. The application of a high magnetic field causes a metamagnetic transition. In a magnetic field applied perpendicular to the polar axis, a helix-to-canted antiferromagnetic transition takes place through the intermediate soliton lattice type state. On the other hand, a magnetic field applied along the polar axis induces a first-order metamagnetic transition. These metamagnetic transitions accompany a change in the electric polarization along the polar axis.

\end{abstract}

\maketitle

\section{Introduction}
Noncentrosymmetric magnets often host non-collinear or non-coplanar spin arrangements, such as magnetic helices, solitons, and skyrmions~\cite{dzyaloshinsky1958thermodynamic, moriya1960anisotropic}. For instance, magnetic helix, cone, Bloch-type-skyrmion lattice, and chiral soliton lattice have been reported in B20-type compounds~\cite{muhlbauer2009skyrmion}, $\beta$-Mn-type Co-Zn-Mn alloys~\cite{tokunaga2015new}, and Cu$_2$OSeO$_3$~\cite{seki2012observation} with chiral cubic structures. N\'eel-type-skyrmion lattice is observed in polar magnets, such as GaV$_4$S$_8$~\cite{kezsmarki2015neel}, GaV$_4$Se$_8$~\cite{fujima2017thermodynamically}, and VOSe$_2$O$_5$~\cite{kurumaji2017neel}. Furthermore, antiskyrmions appear in magnets with $D_{2d}$-symmetry~\cite{nayak2017magnetic}. The skyrmion lattices in insulating materials, such as Cu$_2$OSeO$_3$ and GaV$_4$S$_8$ have been found to accompany the magnetoelectric (ME) coupling~\cite{seki2012magnetoelectric,ruff2015multiferroicity}. The cross-correlation response attracts interest in terms of the electric-field control of magnetic structure, in particular topological magnetic objects.
\par
To explore a colossal ME response with specific spin ordering, magnetic oxides of \nto-type chiral polar ordered corundum structure  with a space group $R$3 may be good candidates~\cite{cai2017polar}. \nto\ undergoes an antiferromagnetic transition at 52\,K~\cite{vzivkovic2010ni3teo6}. Below the N\'eel temperature, the material shows colossal ME effects across two-step spin-flop transitions~\cite{oh2014non,kim2015successive,yokosuk2016magnetoelectric}. The present study  focuses on isostructural \niso, which is obtained by substitution of In$^{3+}$ and Sb$^{5+}$ for one third of Ni$^{2+}$ and Te$^{6+}$, respectively. The lattice parameters are $a = 5.2168\,{\rm\AA}$ and $c = 14.0166\,{\rm\AA}$ in the hexagonal notation (we use in the hexagonal notation in this paper). While Ni moments in \nto\ are collinearly arranged in the antiferromagnetic phase, \niso\ hosts an incommensurate helimagnetic modulation with a propagation vector \qvec\  = (0, 0.029, 0) below $T_{\rm N}$ = 76\,K, according to the powder neutron diffraction~\cite{ivanov2013spin}. The long helimagnetic period suggests that Dzyaloshinskii-Moriya (DM) interaction should be essential for the helimagnetic order. The helimagnetically ordered Ni planes are stacked along the $c$-axis in the out-of-phase manner as in Fe$_3$PO$_4$O$_3$ ~\cite{ross2015nanosized}. Due to the noncentrosymmetric nature of the underlying crystal structure with both chirality and polarity, \niso\ can exhibit a unique magnetic property and fascinating ME responses. As reported in Ref.~\cite{bogdanov1989thermodynamically}, DM interaction activated in a $C_3$-symmetry magnet can work differently from that in polar or chiral magnets. \niso\ is a rare example which belongs to $C_3$ point group and hosts non-collinear spiral spin ordering. 
\par
Here, we study physical properties of \niso\ by using single crystals. Since a spin helix shows an anisotropic response to an external field in general, experimental research by using single crystalline samples is essential to clarify the multiferroic property of the unique spin spiral order~\cite{2003magnetic}. We have found that large pyroelectricity is induced by the helimagnetic order. Metamagnetic transitions are observed by measurements of magnetization, electric polarization, and a dielectric constant in high magnetic fields. 
\par

\section{Experimental}

Single crystals of \niso\ were grown by the chemical vapor transport method with use of PtCl$_2$ as the transport agent~\cite{weil2014crystal}. The $c$-plane was easily grown, resulting in plate-shaped crystals. The typical dimensions of obtained crystals were 1-3\,mm$^2$ in area and 500\,\textmu m thick. Polarized light microscopy clarified that they had chiral twins, as reported in~\cite{prosnikov2019lattice}. Since it was reported that \nto, which has a similar structure, hosted composite domain walls of chirality and polarity~\cite{wang2015interlocked}, the measurement of electric polarization along the $c$-axis was carried out on a homochiral domain, whose area was 0.74\,mm$^2$ by the integration of displacement current measured by an electrometer (6517A, Keithley). Dielectric constants were measured by using an LCR meter (E4980A, Agilent). Magnetization was measured by a superconducting quantum interference device magnetometer (MPMS-XL, Quantum Design). High magnetic-field measurements of magnetization and electric polarization were performed with use of a pulse magnet at the Institute for Solid State Physics, the University of Tokyo. Dielectric constant measurements in steady high magnetic fields were performed at the High Field Laboratory for Superconducting Materials, Institute for Materials Research, Tohoku University. Small-angle soft x-ray scattering (SAXS) measurements at Ni $L_3$ absorption edge were performed on BL-16A and 19B, Photon Factory, KEK, Japan. The experimental setup for the SAXS measurements is schematically shown in Fig.\,\ref{fig3}(a) (see Refs.~\cite{yamasaki2013diffractometer, yamasaki2015dynamical} for details). A plate of \niso\ crystal of a thickness of about 300\,nm  was fabricated by using a focused ion beam thinning method. The thin-plate sample was put on a Si$_3$N$_4$ membrane covered by a gold film with a pin hole of 5\,\textmu m in diameter as depicted in Fig.\,\ref{fig3}(b). We used left-handed circularly polarized and unpolarized x-ray at BL-16A and BL-19B, respectively.  853-eV (Ni $L_3$ edge) soft x-ray was irradiated on the sample and the scattered x-ray near the transmitted beam was recorded by a charge-coupled detector (CCD) camera. A time-of-flight neutron scattering measurement was carried out on three crystals aligned on an Al plate on BL15 in MLF, J-PARC, Japan~\cite{takata2015design}. Small- and wide-angle neutron scattering were recorded by position sensitive detectors.

\begin{figure}
\centering
\includegraphics[width=9cm]{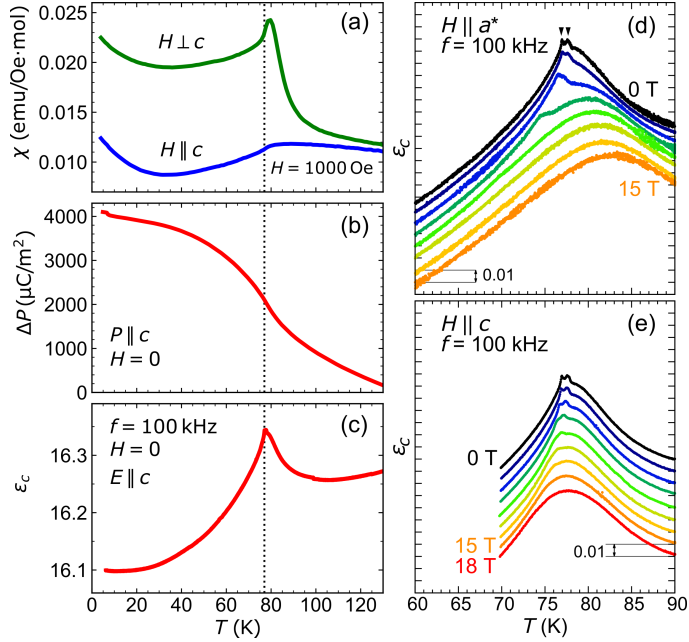}
\caption{Temperature dependence of (a) magnetic susceptibility $\chi$, (b) change in electric polarization $\Delta P$, and (c) electric permittivity $\epsilon_c$ along the $c$-axis at a frequency 100 kHz. A dotted line shows a magnetic transition at 77\,K. (d)(e) Temperature dependence of $\epsilon_c$ in magnetic fields of 0\,T, 2\,T, 4\,T, 6\,T, 8\,T, 10\,T, 12\,T, 15\,T for (d) \Hparaa\ and (e) \Hparac. 18-T data are also added for \Hparac. Data are vertically offset for clarity.}
\label{fig1}
\end{figure}

\section{Fundamental physical properties}

Figure\,\ref{fig1} shows fundamental physical properties of \niso\ as functions of temperature $T$. Magnetic susceptibility shows an anomaly at $T_{\rm N}$ = 77\,K. The temperature approximately agrees with the previously reported magnetic transition temperature~\cite{ivanov2013spin}. Weiss temperature $\theta_{\rm W}$ is estimated to be $-$207\,K and $-$188\,K by using the susceptibility data above 150\,K (not shown) for magnetic fields \Hperpc\ and \Hparac\, respectively. The frustration parameter $|\theta_{\rm W}/T_{\rm N}|$ is smaller than 3, implying that the helimagnetic order should not originate from magnetic frustration but from DM interaction. Figure\,\ref{fig1}(b) describes that the electric polarization shows a steep change at around $T_{\rm N}$ superposed on large pyroelectricity purely due to the polar nature of the crystal. The change in the electric polarization between $T_{\rm N}$ and the lowest temperature is approximately 2000 \textmu${\rm C/m^2}$. This value is almost comparable to \nto~\cite{oh2014non} and larger than the typical value of the order of \textmu${\rm C/m^2}$ in other multiferroic materials~\cite{tokura2014multiferroics}. 

Figures \ref{fig1}(d) and (e) display temperature dependence of electric permittivity $\epsilon_c$ along the $c$-axis around $T_{\rm N}$ in various magnetic fields. A double peak structure is observed around $T_{\rm N}$ in zero field, as indicated by black triangles, which suggests two-step successive phase transitions. An in-plane magnetic field \Hparaa\ shifts the low-temperature anomaly to lower temperatures. The anomaly becomes less prominent with increasing the magnetic field. In contrast, a magnetic field along the $c$-axis only broadens the anomaly, as shown in Fig.\,\ref{fig1}(e).
\par

\section{Small-angle resonant soft x-ray magnetic scattering}

\begin{figure}
\centering
\includegraphics*[width=9cm]{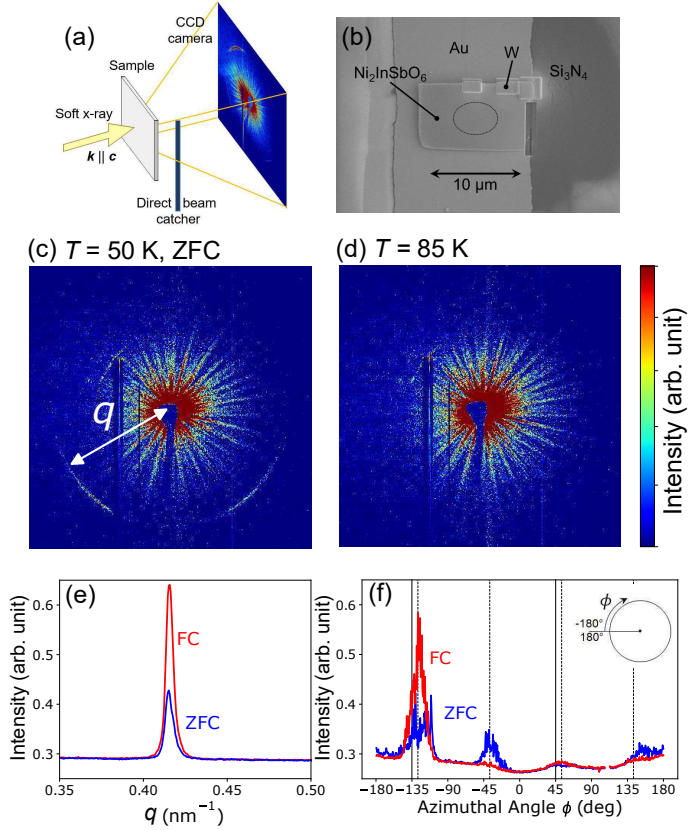}
\caption{Small angle resonant soft x-ray scattering of \niso\ for the incident x-ray propagating parallel to the $c$-axis. (a) Schematic image of experimental setup for SAXS measurements. (b) Thin plate fabricated by a  focused ion beam process on a gold-coated Si$_3$N$_4$ membrane. The doted line indicates a position of the precise circle shaped pinhole.  (c)(d) Diffraction patterns in zero magnetic field at (c) 50\,K (below $T_{\rm N}$) and (d) 85\,K (above $T_{\rm N}$). (e) Intensity profiles of x-ray scattering along the radial direction \qvec\ at 50\,K. The intensity is obtained by the integration in the range of $-142.5^\circ < \phi < -112.5^\circ$. (f) Intensity profiles of x-ray scattering along the azimuthal direction at 50\,K for $0.406\,{\rm nm}^{-1} < |\qvec| < 0.423\,{\rm nm}^{-1}$. The inset describes the definition of azimuthal angle $\phi$. Two Solid lines show magnetic field direction, which is positioned at $\phi = -135^\circ$ and $45^\circ$.}
\label{fig3} 
\end{figure}

A SAXS measurement in resonant with Ni $L_3$ absorption edge was performed with the incident x-ray propagating in the \cvec-direction. Ring-like scattering in the $c$-plane is observed at 50\,K (below $T_{\rm N}$) after zero-field cooling, as shown in Fig.\,\ref{fig3}(c). Note that the shadow of a direct beam catcher is seen at the center of the image. The diffraction disappears at 85\, K (above $T_{\rm N}$), as shown in Fig.\,\ref{fig3}(d), suggesting its magnetic origin. 
Figure\,\ref{fig3}(e) shows the intensity profiles along $|\qvec|$ at an azimuthal angle $\phi = -127.5^\circ$. By pseudo-Voigt fitting of the profile along the radial direction after zero-field cooling , the position and the half width of half maximum of the superlattice peak are estimated to be 4.15$\times 10^{-1}$\,nm$^{-1}$ and 5.73$\times 10^{-3}$\,nm$^{-1}$, respectively. The sharp peak profile indicates the well-defined period $\lambda = 15 \pm 1$\,nm of the helimagnetic order, which is in accord with the previous report\cite{ivanov2013spin}. Four-fold like intensity distribution is observed in the azimuthal dependence of the zero-field cooling pattern, as shown in Figs.\,\ref{fig3}(c) and \ref{fig3}(f), although \niso\ has the $C_3$ symmetry. This discrepancy may arise from some strain in the thin-plate sample and the boundary condition of the square-shaped sample, shown in Fig.\,\ref{fig3}(b). The modulation direction can be controlled by a magnetic-field cooling. Figure\,\ref{fig3}(f) compares the intensity profile along the azimuthal angle $\phi$ in zero magnetic field  at 50\,K after zero magnetic field cooling with that after the sample was cooled from above $T_{\rm N}$ in a magnetic field $\mu_0H = 0.4$\,T in the direction of $\phi = 45^\circ$. The scattering intensities around $\phi \simeq -135^\circ$ and 45$^\circ$, which locate along the field direction, are increased after the field-cooling procedure, while those around $\phi \simeq -45^\circ$ and $135^\circ$ are decreased. Considering the fact that a magnetic field tends to rotate the propagation direction of the helix so that the spiral plane becomes normal to the magnetic field direction to maximize the Zeeman-energy gain, the observed field-cooling effect implies that the helimagnetic order should be nearly proper-screw type. The chiral component of DM interaction (DM vector parallel to the bond) should be dominant over the polar one (DM vector normal to the bond).

\section{Neutron diffraction}

\begin{figure}
\centering
\includegraphics[width=8.5cm]{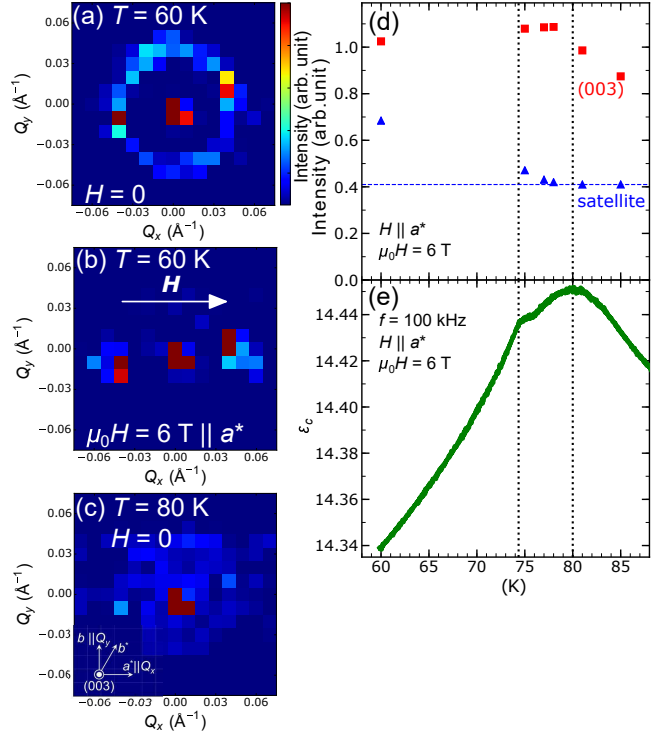}
\caption{Neutron scattering patterns in the $(hk3)$ plane. $Q_x$ and $Q_y$ are along $a^*$- and $b$-axes, respectively. The inset in (c) shows the reciprocal crystal axes in the experimental setup. (a) Scattering pattern at 0\,T and 60\,K. (b) Scattering pattern at $\mu_0H = 6$\,T and 60\,K. The magnetic field was applied along the $a^*$-axis before the sample was cooled down to the ordered phase. (c) Scattering pattern at 80\,K (above $T_{\rm N}$) (d) The temperature dependence of the intensity of (003) and satellite diffractions described as red squares and blue triangles, respectively. The background level of satellite diffractions is depicted as a blue horizontal broken line determined by the scattering intensity at 85\,K (far above $T_{\rm N}$). (e) The temperature dependence of dielectric constant in a magnetic field $\mu_0H$ of 6\,T along the $a^*$-axis. The vertical dot lines indicate the temperature where dielectric anomalies appear.}
\label{fig4}
\end{figure}

The effect of a magnetic field on the helicoid is also confirmed by neutron scattering. Figure\,\ref{fig4} illustrates a contour map of the intensity of neutron diffraction on the two-dimensional reciprocal $(hk3)$ plane. Ring-like scattering is observed around (003) reflection at 60\,K (below $T_{\rm N}$) and disappears at 80\,K (above $T_{\rm N}$). The ring shape indicates that the direction of the modulation vector \mbox{\boldmath $q$} of helimagnetic order in the bulk sample should be distributed isotropically in the $c$-plane.  The period of magnetic modulation is estimated to be about 15.7\,nm, which agrees with the previous report~\cite{ivanov2013spin} as well as the present SAXS result. The application of a magnetic field of 6\,T along the $a^*$-axis at 60\,K concentrates the intensity of the magnetic diffraction along the field (Fig.\,\ref{fig4}(b)), which is again in accord with the SAXS result. Figure\,\ref{fig4}(d) shows the temperature dependence of (003)  and satellite intensities at 6\,T. The (003) intensity reaches the maximum just below $T_{\rm N}$, and decreases above 80\,K. On the other hand, the satellite intensity is decreased with a rise of temperature, and is almost the same as the background level above 80\,K. These behaviors well correspond to the anomalies of dielectric constant in a magnetic field of 6\,T, as depicted in Fig.\,\ref{fig4}(e). The system may first undergo a magnetic transition to the commensurate layered antiferromagnetic phase upon cooling and then successively enter the helical phase with a long-wavelength modulation below 74\,K.
\par

\section{Pulse magnet measurement}

\begin{figure}
\centering
\includegraphics[width=7.5cm]{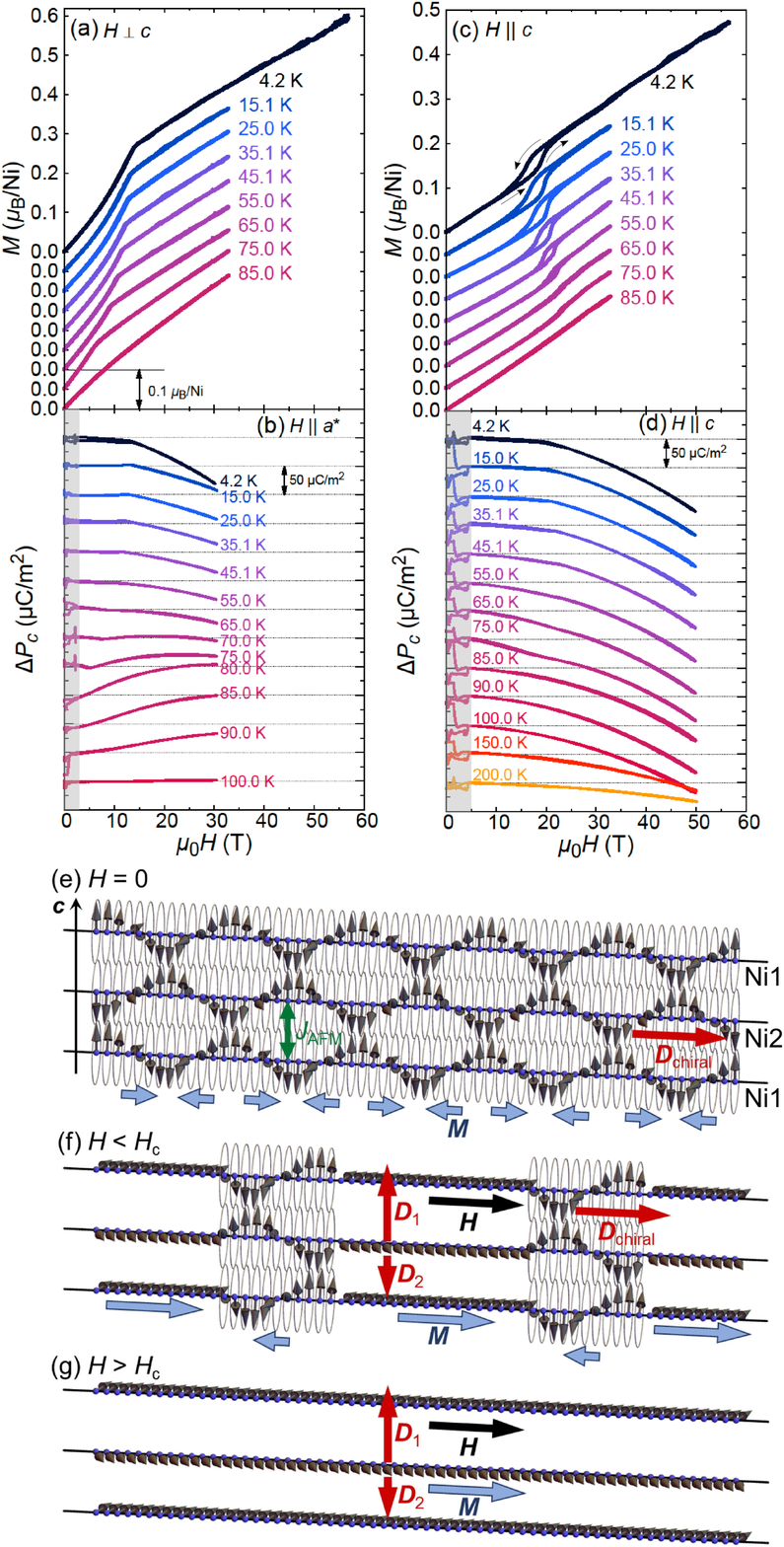}
\caption{(a)(c) Magnetization curves at various temperatures in (a) \Hperpc\ and (c) \Hparac\ configurations. (b)(d) Magnetic-field dependence of the change in electric polarization $\Delta P_c$ along the $c$-axis at various temperatures in (b) \Hparaa\ and (d) \Hparac\ configurations. Anomalies in the shaded area at low fields are due to electric noise of the ignition of the pulse magnet. $M$-$H$ and $P$-$H$ curves in (a)-(d) are vertically offset for clarity. (e)-(g) Schematic drawing of the possible magnetization process in \Hperpc. For simplicity, polarity-induced uniform DM interaction is neglected. (e) Proper-screw helicoid in zero field. (f) Intermediate soliton lattice state induced by a moderate magnetic field $H < H_c$. (g) Weak-ferromagnetic structure in a magnetic field higher than the critical field $H_{\rm c}$.}
\label{fig2}
\end{figure}

Figure \ref{fig2} shows high-field $M$-$H$ and $P$-$H$ curves at various temperatures measured by using a pulse magnet. For \Hperpc, the $M$-$H$ curves below $T_{\rm N}$ have a clear anomaly, as shown in Fig.\,\ref{fig2}(a). The transition field $H_c$ is approximately 14\,T at 4.2\,K, decreases monotonically with a rise in temperature, and disappears above $T_{\rm N}$, as in the curve at 85\,K. The magnetization below $H_c$ is superlinear to the field, which could be featured by soliton lattice formation, as discussed later. The magnetization in a higher-field phase below $T_{\rm N}$ is linear to the magnetic field, and the extrapolation to $\mu_0H = 0$\,T remains non-zero (0.17\,$\mu_B$/Ni at 4.2\,K), suggesting that the higher-field phase should essentially have a spontaneous magnetization component. The extrapolation value decreases monotonically with an increase of temperature. 
For \Hparac, a metamagnetic transition takes place with a hysteresis loop, which is completely different from the case of \Hperpc\ configuration, as shown in Fig.\,\ref{fig2}(c). The transition field increases and the hysteresis loop becomes smaller with a rise of temperature. As in the case of \Hperpc\ configuration, the field-induced magnetic transition disappears above $T_{\rm N}$.

As shown in Figs.\,4(b) and (d), the change in electric polarization shows an anomaly at the metamagnetic transition. The electric polarization below $T_{\rm N}$ is almost insensitive to the magnetic field perpendicular to the $c$-axis in the low-field phase, while in the high-field phase, it shows parabolic dependence on magnetic field. Around $T_{\rm N}$, the sign of the ME coefficient is reversed, and the parabolic dependence disappears. The value of $\Delta P_c$ at each temperature in the high field phase is in the order of 10 \textmu C/m$^2$, which is much smaller than the pyroelectricity at zero-field. The ME effect remains even above $T_{\rm N}$. The electric polarization shows quadratic dependence on magnetic field for \Hparac\ in both low and high fields. The typical value is comparable with that in the \Hperpc\ configuration. The quadratic ME effect is observed up to 200\,K. The sign of the ME effect does not change in the whole temperature range, which is different from the case of \Hperpc. 
\par

\section{Magnetic phase diagram}
\begin{figure}
\centering
\includegraphics[width=7cm]{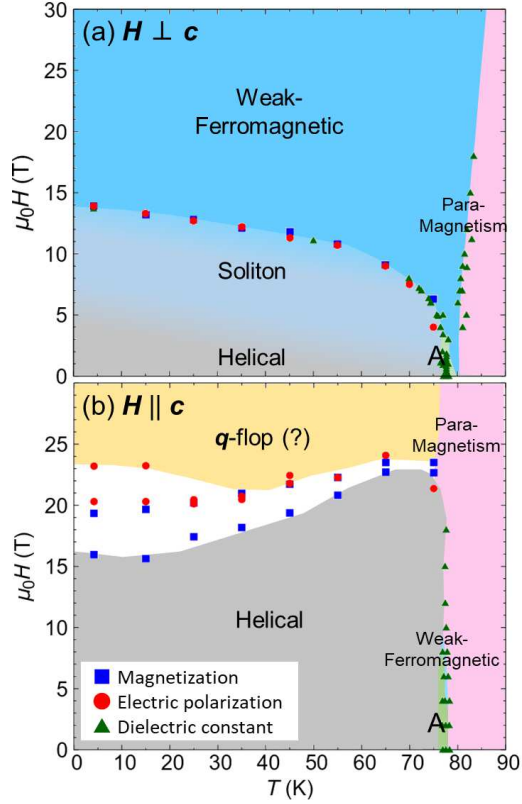}
\caption{Magnetic phase diagrams of \niso\ proposed by macroscopic and quantum beam measurements in configurations (a)\Hperpc\ and (b)\Hparac. A white area which is located between blue and yellow areas in \Hparac\ configuration shows the hysteresis loop (upper and lower points are obtained in the field-increasing and -decreasing processes, respectively).}
\label{fig5}
\end{figure}

Based on the dielectric constant, electric polarization, and magnetization data as well as SAXS and neutron diffraction profiles, we propose magnetic phase diagrams of \niso, as shown in Figs.\,\ref{fig5}\,(a) and (b). The gray and pink areas indicate the helical and paramagnetic phases, respectively, as previously reported in~\cite{ivanov2013spin}. A small green pocket ``A" between helical and paramagnetic phases is revealed by the dielectric property. The nature of this phase is still unknown, because SAXS and neutron diffraction intensities are too weak to analyze. The phases displayed by blue and yellow areas are discovered in the present high-field measurements.     
\par

\section{Discussions}

We discuss the variations in the magnetic structure of \niso. We assign the low-field phase in the \Hperpc\ configuration below $H_c$ to the soliton lattice phase with modulating the nearly proper-screw type helicoid (Fig.\,\ref{fig2}(e)), as signaled in the superlinear $M$-$H$ curve (Fig.\,\ref{fig2}(a)).  As shown in Fig.\,\ref{fig4}(b), the \qvec-direction is aligned nearly along the external field in \niso\ below several tesla. With increasing the magnetic field, the helical structure is gradually modified in the \Hparaq\ configuration. In a chiral ferromagnet with uniaxial \qvec, chiral-soliton lattice can be formed in the \Hperpq\ configuration~\cite{pappas2012new}. On the other hand, the formation of solitons in a magnetic field is not straightforward in a chiral antiferromagnet. Above a critical field $H_{\rm c}$, weak-ferromagnetic structure should appear as is the case in BiFeO$_3$~\cite{ederer2005weak, tokunaga2015magnetic}. 
To explain the emergence of the soliton and weak-ferromagnetic phase in the \Hperpc\ configuration, we take into account an additional alternate DM interaction, because neither chirality- nor polarity-induced uniform DM interaction can drive the spontaneous magnetization.
In the present system, layers formed by crystallographically non-equivalent two Ni sites, namely Ni1 and Ni2, stack alternately along the $c$-axis. Here we consider the interaction between Ni moments on neighboring layers. In addition to the antiferromagnetic symmetric exchange, the lack of inversion center allows antisymmetric exchange interactions. The essential component of the DM vector is along the $c$-axis, which should alternately change the value. Here we show alternating component,  represented by DM vectors \mbox{\boldmath $D$}$_1$ and \mbox{\boldmath $D$}$_2$ along the $c$-axis with the opposite signs to each other, as shown in Fig.\,\ref{fig2}(g)~\cite{DMI}. The DM vectors make magnetic moments on Ni1 and Ni2 noncollinear if Ni moments are in the $c$-plane, while the antisymmetric exchange is inactive if Ni moments are along the $c$-axis. Therefore, when the Ni moments are arranged to form layered antiferromagnetic structure, a uniform canted weak-ferromagnetic component can emerge in the $c$-plane. This situation is expected to realize above the critical field $H_c$, as shown in Fig.\,\ref{fig2}(g). Here, the weak-ferromagnetic component roughly orients in the \Hvec-direction. When the magnetic field is reversed, the weak-ferromagnetic moment flips, which should accompany the reversal of the sublattice moment. At $H = 0$, on the other hand, the staggered DM vectors induce local weak-ferromagnetic moments at positions where Ni moments have the $c$-plane components on the proper-screw structure. These local weak-ferromagnetic components are along the \qvec-direction, and bring about sinusoidal modulation in the original proper-screw structure (Fig.\,\ref{fig2}(e)). Hence, the application of a magnetic field in the $c$-plane modulates the sinusoid (and thus the underlying proper-screw structure) so that the regions which have local weak-ferromagnetic moment parallel (antiparallel) to the field expand (shrink) to acquire Zeeman energy gain, resulting in the formation of the soliton structure, as shown in Fig.\,\ref{fig2}(f).
\par
On the contrary, the isothermal magnetization curve in the \Hparac\ configuration is almost linear below the critical field. The difference between two magnetic field configurations can be caused whether the local weak-ferromagnetic component is stabilized by DM interaction. The nature of the observed field-induced magnetic transition in \Hparac\ configuration is not clear in the present stage. Conical structure with the modulation along the $c$-axis may appear in a high external magnetic field along $c$-axis as in Co doped Ni$_2$ScSbO$_6$~\cite{ji2018lock}. Another possible scenario is that a simple canted layered antiferromagnetic structure is induced similarly to the case of \Hperpc.

The temperature dependence of electric polarization is shown in Fig.\,\ref{fig1}(b). The proper-screw type helicoid drives considerably large pyroelectricity. 
To consider the origin of the ME properties, we take inverse DM mechanism~\cite{katsura2005spin}, exchange striction, and spin-direction dependent $p$-$d$ hybridization~\cite{arima2007ferroelectricity} into account. The inverse DM effect cannot work effectively in proper-screw helicoid. The electric polarization driven by the exchange striction mechanism $\Delta P^{\rm{ex}}_z$ is expressed as
\begin{align}
\Delta P^{\rm{ex}}_z &= C\mbox{\boldmath $S$}_{\rm Ni1}\cdot \mbox{\boldmath $S$}_{\rm Ni2}.
\end{align}
The pyroelectricity measurement shows that the antiferromagnetic arrangement between $\mbox{\boldmath $S$}_{\rm Ni1}$ and $\mbox{\boldmath $S$}_{\rm Ni2}$ enhances the polarization. Therefore the coefficient $C$ should be negative if the exchange striction is dominant. In the proposed weak-ferromagnetic structure as described in Fig.\,\ref{fig2}(g), Ni1 and Ni2 moments can be represented as $\mbox{\boldmath $S$}_{\rm Ni1} = (S_x, S_y, 0)$ and $\mbox{\boldmath $S$}_{\rm Ni2} = (S_x, -S_y, 0)$, where $x$ is the magnetic-field direction and $y$ is set in the $c$-plane. The exchange striction is calculated to be $C(-S^2 + 2S_x^2)$, where $S$ is the value of Ni spin moment.
\par
To calculate the contribution of $p$-$d$ hybridization mechanism to the electric polarization in a NiO$_6$ octahedral cluster with 3-fold rotational symmetry, we set six unit vectors $\mbox{\boldmath $e$}_{ij}^{\rm{NiO}}$ along the Ni-O bonds as
\begin{align}
\mbox{\boldmath $e$}_{11}^{\rm{NiO}} &=\nonumber 
\begin{pmatrix}
\sin\theta_1 \cos\phi_1\\
\sin\theta_1 \sin\phi_1\\
\cos\theta_1
\end{pmatrix},\,
\mbox{\boldmath $e$}_{12}^{\rm{NiO}} = 
\begin{pmatrix}
\sin\theta_1 \cos\left(\phi_1 + \frac{2}{3}\pi\right)\\
\sin\theta_1 \sin\left(\phi_1 + \frac{2}{3}\pi\right)\\
\cos\theta_1
\end{pmatrix},\\
\mbox{\boldmath $e$}_{13}^{\rm{NiO}} &=\nonumber 
\begin{pmatrix}
\sin\theta_1 \cos\left(\phi_1 - \frac{2}{3}\pi\right)\\
\sin\theta_1 \sin\left(\phi_1 - \frac{2}{3}\pi\right)\\
\cos\theta_1
\end{pmatrix},\,
\mbox{\boldmath $e$}_{21}^{\rm{NiO}} = 
\begin{pmatrix}
\sin\theta_2 \cos\phi_2\\
\sin\theta_2 \sin\phi_2\\
\cos\theta_2
\end{pmatrix},\\
\mbox{\boldmath $e$}_{22}^{\rm{NiO}} &=\nonumber 
\begin{pmatrix}
\sin\theta_2 \cos\left(\phi_2 + \frac{2}{3}\pi\right)\\
\sin\theta_2 \sin\left(\phi_2 + \frac{2}{3}\pi\right)\\
\cos\theta_2
\end{pmatrix},\\
\mbox{\boldmath $e$}_{23}^{\rm{NiO}} &= 
\begin{pmatrix}
\sin\theta_2 \cos\left(\phi_2 - \frac{2}{3}\pi\right)\\
\sin\theta_2 \sin\left(\phi_2 - \frac{2}{3}\pi\right)\\
\cos\theta_2
\end{pmatrix},
\end{align}
where the index $i$ denotes whether the oxygen positions upward ($i = 1$) or downward ($i = 2$) Ni. The index $j$ denotes 3-fold symmetric Ni-O bonds in the $c$-plane. The values of $\phi_1, \phi_2, \theta_1$ and $\theta_2$ are $65.9^\circ, 4.5^\circ, 48.7^\circ$, and $115.9^\circ$ for the Ni1-O cluster and $1.5^\circ, 54.1^\circ, 49.2^\circ$, and $120.6^\circ$ for the Ni2-O cluster, respectively~\cite{ivanov2013spin}. The change in the electric polarization induced by spin-direction dependent $p$-$d$ hybridization mechanism $\Delta P_z^{pd}$ is given by
\begin{align}
\Delta P_z^{pd} &\propto \sum_{i, j} (\mbox{\boldmath $S$}_{\rm{Ni}}\cdot\mbox{\boldmath $e$}_{ij}^{\rm{NiO}})^2e^{\rm{NiO}}_{ijz} \\
&= A(S_x^2 + S_y^2)+ B S_z^2\\
&= AS^2 + (B - A)S_z^2,
\end{align}
where $A$ and $B$ are constants. This formula indicates that only the $z$-component of magnetic moment can affect $P_z$. 
The $z$-component of magnetization fluctuates around $T_{\rm N}$. The $P_z$ value should change when the external magnetic field suppresses the fluctuation and modifies $\langle S_z^2\rangle$. The sign of $\Delta P^{\rm{ex}}_z$ is negative both for \Hparac\ and for \Hperpc\ while that of $\Delta P^{pd}_z$ is positive for \Hperpc\ but negative for \Hparac.
The field dependence of electric polarization is shown in Fig.\,\ref{fig2}(b) at 70-90\,K. At lower temperatures, it seems that the change in electric polarization is dominated by exchange striction mechanism, because $S_z = 0$ and thus $\Delta P_z^{pd} \simeq 0$ in the field-induced weak-ferromagnetic phase. On the other hand, for \Hparac, $S_z$ is modified as $H$ is increased and thus $\Delta P_z^{pd} \neq 0$. The magnetization is approximately proportional to the field in each magnetic phase. Therefore, the electric polarization driven by $p$-$d$ hybridization should show quadratic $H$-dependence as in the case of the exchange striction, which agrees with the experimental observation. 
\par


\section{conclusion}

We have found giant pyroelectricity in \niso\ driven by helimagnetic order. A peculiar magnetization curve in the \Hperpc\ configuration suggests soliton-like magnetic modulation. We propose that 2$\pi$ solitons may be created in the canted antiferromagnetic background. The application of a magnetic field along $c$-axis triggers a metamagnetic transition with a hysteresis loop. Further neutron scattering study is necessary to pin down the magnetic structure at each phase.
\par

\section{acknowledgements}

This work was partly supported by JSPS KAKENHI Grants No.s\,JP25220803, JP16H01065, JP16K13828, JP19H01835, and JP19H05826. Y.A., T.S. and Y.F. are supported by Japan Society for the Promotion of Science (JSPS) through Program for Leading Graduate Schools(MERIT). The measurements of magnetization, electric polarization, and x-ray Laue photograph were performed by using facilities of the Institute for Solid State Physics (ISSP), the University of Tokyo. The authors thank T. Yajima for assistance in the XRD experiment at ISSP. Measurements in steady high magnetic fields were performed in the High Field Laboratory for Superconducting Materials, Institute for Materials Research, Tohoku University, Japan (Project No.\,16H0023). The synchrotron soft x-ray measurements were performed under User Programs No.\,2015S2-007 at BL-16A and No.\,2016PF-BL-19B at BL-19B, Photon Factory, KEK, Japan. The neutron scattering measurement were performed under User Programs No.\,2017A0069 at the Materials and Life Science Experimental Facilities, J-PARC, Japan. 



\begin{thebibliography}{34}
\expandafter\ifx\csname natexlab\endcsname\relax\def\natexlab#1{#1}\fi
\expandafter\ifx\csname bibnamefont\endcsname\relax
  \def\bibnamefont#1{#1}\fi
\expandafter\ifx\csname bibfnamefont\endcsname\relax
  \def\bibfnamefont#1{#1}\fi
\expandafter\ifx\csname citenamefont\endcsname\relax
  \def\citenamefont#1{#1}\fi
\expandafter\ifx\csname url\endcsname\relax
  \def\url#1{\texttt{#1}}\fi
\expandafter\ifx\csname urlprefix\endcsname\relax\def\urlprefix{URL }\fi
\providecommand{\bibinfo}[2]{#2}
\providecommand{\eprint}[2][]{\url{#2}}

\bibitem[{\citenamefont{Dzyaloshinsky}(1958)}]{dzyaloshinsky1958thermodynamic}
\bibinfo{author}{\bibfnamefont{I.}~\bibnamefont{Dzyaloshinsky}},
  \bibinfo{journal}{J. Phys. Chem. Solids} \textbf{\bibinfo{volume}{4}},
  \bibinfo{pages}{241} (\bibinfo{year}{1958}).

\bibitem[{\citenamefont{Moriya}(1960)}]{moriya1960anisotropic}
\bibinfo{author}{\bibfnamefont{T.}~\bibnamefont{Moriya}},
  \bibinfo{journal}{Phys. Rev.} \textbf{\bibinfo{volume}{120}},
  \bibinfo{pages}{91} (\bibinfo{year}{1960}).

\bibitem[{\citenamefont{M{\"u}hlbauer et~al.}(2009)\citenamefont{M{\"u}hlbauer,
  Binz, Jonietz, Pfleiderer, Rosch, Neubauer, Georgii, and
  B{\"o}ni}}]{muhlbauer2009skyrmion}
\bibinfo{author}{\bibfnamefont{S.}~\bibnamefont{M{\"u}hlbauer}},
  \bibinfo{author}{\bibfnamefont{B.}~\bibnamefont{Binz}},
  \bibinfo{author}{\bibfnamefont{F.}~\bibnamefont{Jonietz}},
  \bibinfo{author}{\bibfnamefont{C.}~\bibnamefont{Pfleiderer}},
  \bibinfo{author}{\bibfnamefont{A.}~\bibnamefont{Rosch}},
  \bibinfo{author}{\bibfnamefont{A.}~\bibnamefont{Neubauer}},
  \bibinfo{author}{\bibfnamefont{R.}~\bibnamefont{Georgii}}, \bibnamefont{and}
  \bibinfo{author}{\bibfnamefont{P.}~\bibnamefont{B{\"o}ni}},
  \bibinfo{journal}{Science} \textbf{\bibinfo{volume}{323}},
  \bibinfo{pages}{915} (\bibinfo{year}{2009}).

\bibitem[{\citenamefont{Tokunaga
  et~al.}(2015{\natexlab{a}})\citenamefont{Tokunaga, Yu, White, R{\o}nnow,
  Morikawa, Taguchi, and Tokura}}]{tokunaga2015new}
\bibinfo{author}{\bibfnamefont{Y.}~\bibnamefont{Tokunaga}},
  \bibinfo{author}{\bibfnamefont{X.}~\bibnamefont{Yu}},
  \bibinfo{author}{\bibfnamefont{J.}~\bibnamefont{White}},
  \bibinfo{author}{\bibfnamefont{H.~M.} \bibnamefont{R{\o}nnow}},
  \bibinfo{author}{\bibfnamefont{D.}~\bibnamefont{Morikawa}},
  \bibinfo{author}{\bibfnamefont{Y.}~\bibnamefont{Taguchi}}, \bibnamefont{and}
  \bibinfo{author}{\bibfnamefont{Y.}~\bibnamefont{Tokura}},
  \bibinfo{journal}{Nat. Commun.} \textbf{\bibinfo{volume}{6}},
  \bibinfo{pages}{7638} (\bibinfo{year}{2015}{\natexlab{a}}).

\bibitem[{\citenamefont{Seki et~al.}(2012{\natexlab{a}})\citenamefont{Seki, Yu,
  Ishiwata, and Tokura}}]{seki2012observation}
\bibinfo{author}{\bibfnamefont{S.}~\bibnamefont{Seki}},
  \bibinfo{author}{\bibfnamefont{X.}~\bibnamefont{Yu}},
  \bibinfo{author}{\bibfnamefont{S.}~\bibnamefont{Ishiwata}}, \bibnamefont{and}
  \bibinfo{author}{\bibfnamefont{Y.}~\bibnamefont{Tokura}},
  \bibinfo{journal}{Science} \textbf{\bibinfo{volume}{336}},
  \bibinfo{pages}{198} (\bibinfo{year}{2012}{\natexlab{a}}).

\bibitem[{\citenamefont{K{\'e}zsm{\'a}rki
  et~al.}(2015)\citenamefont{K{\'e}zsm{\'a}rki, Bord{\'a}cs, Milde, Neuber,
  Eng, White, R{\o}nnow, Dewhurst, Mochizuki, Yanai
  et~al.}}]{kezsmarki2015neel}
\bibinfo{author}{\bibfnamefont{I.}~\bibnamefont{K{\'e}zsm{\'a}rki}},
  \bibinfo{author}{\bibfnamefont{S.}~\bibnamefont{Bord{\'a}cs}},
  \bibinfo{author}{\bibfnamefont{P.}~\bibnamefont{Milde}},
  \bibinfo{author}{\bibfnamefont{E.}~\bibnamefont{Neuber}},
  \bibinfo{author}{\bibfnamefont{L.}~\bibnamefont{Eng}},
  \bibinfo{author}{\bibfnamefont{J.}~\bibnamefont{White}},
  \bibinfo{author}{\bibfnamefont{H.~M.} \bibnamefont{R{\o}nnow}},
  \bibinfo{author}{\bibfnamefont{C.}~\bibnamefont{Dewhurst}},
  \bibinfo{author}{\bibfnamefont{M.}~\bibnamefont{Mochizuki}},
  \bibinfo{author}{\bibfnamefont{K.}~\bibnamefont{Yanai}},
  \bibnamefont{et~al.}, \bibinfo{journal}{Nat. Mater.}
  \textbf{\bibinfo{volume}{14}}, \bibinfo{pages}{1116} (\bibinfo{year}{2015}).

\bibitem[{\citenamefont{Fujima et~al.}(2017)\citenamefont{Fujima, Abe,
  Tokunaga, and Arima}}]{fujima2017thermodynamically}
\bibinfo{author}{\bibfnamefont{Y.}~\bibnamefont{Fujima}},
  \bibinfo{author}{\bibfnamefont{N.}~\bibnamefont{Abe}},
  \bibinfo{author}{\bibfnamefont{Y.}~\bibnamefont{Tokunaga}}, \bibnamefont{and}
  \bibinfo{author}{\bibfnamefont{T.}~\bibnamefont{Arima}},
  \bibinfo{journal}{Phys. Rev. B} \textbf{\bibinfo{volume}{95}},
  \bibinfo{pages}{180410} (\bibinfo{year}{2017}).

\bibitem[{\citenamefont{Kurumaji et~al.}(2017)\citenamefont{Kurumaji, Nakajima,
  Ukleev, Feoktystov, Arima, Kakurai, and Tokura}}]{kurumaji2017neel}
\bibinfo{author}{\bibfnamefont{T.}~\bibnamefont{Kurumaji}},
  \bibinfo{author}{\bibfnamefont{T.}~\bibnamefont{Nakajima}},
  \bibinfo{author}{\bibfnamefont{V.}~\bibnamefont{Ukleev}},
  \bibinfo{author}{\bibfnamefont{A.}~\bibnamefont{Feoktystov}},
  \bibinfo{author}{\bibfnamefont{T.}~\bibnamefont{Arima}},
  \bibinfo{author}{\bibfnamefont{K.}~\bibnamefont{Kakurai}}, \bibnamefont{and}
  \bibinfo{author}{\bibfnamefont{Y.}~\bibnamefont{Tokura}},
  \bibinfo{journal}{Phys. Rev. Lett.} \textbf{\bibinfo{volume}{119}},
  \bibinfo{pages}{237201} (\bibinfo{year}{2017}).

\bibitem[{\citenamefont{Nayak et~al.}(2017)\citenamefont{Nayak, Kumar, Ma,
  Werner, Pippel, Sahoo, Damay, R{\"o}{\ss}ler, Felser, and
  Parkin}}]{nayak2017magnetic}
\bibinfo{author}{\bibfnamefont{A.~K.} \bibnamefont{Nayak}},
  \bibinfo{author}{\bibfnamefont{V.}~\bibnamefont{Kumar}},
  \bibinfo{author}{\bibfnamefont{T.}~\bibnamefont{Ma}},
  \bibinfo{author}{\bibfnamefont{P.}~\bibnamefont{Werner}},
  \bibinfo{author}{\bibfnamefont{E.}~\bibnamefont{Pippel}},
  \bibinfo{author}{\bibfnamefont{R.}~\bibnamefont{Sahoo}},
  \bibinfo{author}{\bibfnamefont{F.}~\bibnamefont{Damay}},
  \bibinfo{author}{\bibfnamefont{U.~K.} \bibnamefont{R{\"o}{\ss}ler}},
  \bibinfo{author}{\bibfnamefont{C.}~\bibnamefont{Felser}}, \bibnamefont{and}
  \bibinfo{author}{\bibfnamefont{S.~S.} \bibnamefont{Parkin}},
  \bibinfo{journal}{Nature (London)} \textbf{\bibinfo{volume}{548}},
  \bibinfo{pages}{561} (\bibinfo{year}{2017}).

\bibitem[{\citenamefont{Seki et~al.}(2012{\natexlab{b}})\citenamefont{Seki,
  Ishiwata, and Tokura}}]{seki2012magnetoelectric}
\bibinfo{author}{\bibfnamefont{S.}~\bibnamefont{Seki}},
  \bibinfo{author}{\bibfnamefont{S.}~\bibnamefont{Ishiwata}}, \bibnamefont{and}
  \bibinfo{author}{\bibfnamefont{Y.}~\bibnamefont{Tokura}},
  \bibinfo{journal}{Phys. Rev. B} \textbf{\bibinfo{volume}{86}},
  \bibinfo{pages}{060403} (\bibinfo{year}{2012}{\natexlab{b}}).

\bibitem[{\citenamefont{Ruff et~al.}(2015)\citenamefont{Ruff, Widmann,
  Lunkenheimer, Tsurkan, Bord{\'a}cs, K{\'e}zsm{\'a}rki, and
  Loidl}}]{ruff2015multiferroicity}
\bibinfo{author}{\bibfnamefont{E.}~\bibnamefont{Ruff}},
  \bibinfo{author}{\bibfnamefont{S.}~\bibnamefont{Widmann}},
  \bibinfo{author}{\bibfnamefont{P.}~\bibnamefont{Lunkenheimer}},
  \bibinfo{author}{\bibfnamefont{V.}~\bibnamefont{Tsurkan}},
  \bibinfo{author}{\bibfnamefont{S.}~\bibnamefont{Bord{\'a}cs}},
  \bibinfo{author}{\bibfnamefont{I.}~\bibnamefont{K{\'e}zsm{\'a}rki}},
  \bibnamefont{and} \bibinfo{author}{\bibfnamefont{A.}~\bibnamefont{Loidl}},
  \bibinfo{journal}{Sci. Adv.} \textbf{\bibinfo{volume}{1}},
  \bibinfo{pages}{e1500916} (\bibinfo{year}{2015}).

\bibitem[{\citenamefont{Cai et~al.}(2017)\citenamefont{Cai, Greenblatt, and
  Li}}]{cai2017polar}
\bibinfo{author}{\bibfnamefont{G.-H.} \bibnamefont{Cai}},
  \bibinfo{author}{\bibfnamefont{M.}~\bibnamefont{Greenblatt}},
  \bibnamefont{and} \bibinfo{author}{\bibfnamefont{M.-R.} \bibnamefont{Li}},
  \bibinfo{journal}{Chem. Mater.} \textbf{\bibinfo{volume}{29}},
  \bibinfo{pages}{5447} (\bibinfo{year}{2017}).

\bibitem[{\citenamefont{{\v{Z}}ivkovi{\'c}
  et~al.}(2010)\citenamefont{{\v{Z}}ivkovi{\'c}, Pr{\v{s}}a, Zaharko, and
  Berger}}]{vzivkovic2010ni3teo6}
\bibinfo{author}{\bibfnamefont{I.}~\bibnamefont{{\v{Z}}ivkovi{\'c}}},
  \bibinfo{author}{\bibfnamefont{K.}~\bibnamefont{Pr{\v{s}}a}},
  \bibinfo{author}{\bibfnamefont{O.}~\bibnamefont{Zaharko}}, \bibnamefont{and}
  \bibinfo{author}{\bibfnamefont{H.}~\bibnamefont{Berger}},
  \bibinfo{journal}{J. Phys.: Condens. Matter} \textbf{\bibinfo{volume}{22}},
  \bibinfo{pages}{056002} (\bibinfo{year}{2010}).

\bibitem[{\citenamefont{Oh et~al.}(2014)\citenamefont{Oh, Artyukhin, Yang,
  Zapf, Kim, Vanderbilt, and Cheong}}]{oh2014non}
\bibinfo{author}{\bibfnamefont{Y.~S.} \bibnamefont{Oh}},
  \bibinfo{author}{\bibfnamefont{S.}~\bibnamefont{Artyukhin}},
  \bibinfo{author}{\bibfnamefont{J.~J.} \bibnamefont{Yang}},
  \bibinfo{author}{\bibfnamefont{V.}~\bibnamefont{Zapf}},
  \bibinfo{author}{\bibfnamefont{J.~W.} \bibnamefont{Kim}},
  \bibinfo{author}{\bibfnamefont{D.}~\bibnamefont{Vanderbilt}},
  \bibnamefont{and} \bibinfo{author}{\bibfnamefont{S.-W.}
  \bibnamefont{Cheong}}, \bibinfo{journal}{Nat. Commun.}
  \textbf{\bibinfo{volume}{5}}, \bibinfo{pages}{3201} (\bibinfo{year}{2014}).

\bibitem[{\citenamefont{Kim et~al.}(2015)\citenamefont{Kim, Artyukhin, Mun,
  Jaime, Harrison, Hansen, Yang, Oh, Vanderbilt, Zapf
  et~al.}}]{kim2015successive}
\bibinfo{author}{\bibfnamefont{J.~W.} \bibnamefont{Kim}},
  \bibinfo{author}{\bibfnamefont{S.}~\bibnamefont{Artyukhin}},
  \bibinfo{author}{\bibfnamefont{E.~D.} \bibnamefont{Mun}},
  \bibinfo{author}{\bibfnamefont{M.}~\bibnamefont{Jaime}},
  \bibinfo{author}{\bibfnamefont{N.}~\bibnamefont{Harrison}},
  \bibinfo{author}{\bibfnamefont{A.}~\bibnamefont{Hansen}},
  \bibinfo{author}{\bibfnamefont{J.}~\bibnamefont{Yang}},
  \bibinfo{author}{\bibfnamefont{Y.~S.} \bibnamefont{Oh}},
  \bibinfo{author}{\bibfnamefont{D.}~\bibnamefont{Vanderbilt}},
  \bibinfo{author}{\bibfnamefont{V.~S.} \bibnamefont{Zapf}},
  \bibnamefont{et~al.}, \bibinfo{journal}{Phys. Rev. Lett.}
  \textbf{\bibinfo{volume}{115}}, \bibinfo{pages}{137201}
  (\bibinfo{year}{2015}).

\bibitem[{\citenamefont{Yokosuk et~al.}(2016)\citenamefont{Yokosuk, Al-Wahish,
  Artyukhin, O’Neal, Mazumdar, Chen, Yang, Oh, McGill, Haule
  et~al.}}]{yokosuk2016magnetoelectric}
\bibinfo{author}{\bibfnamefont{M.}~\bibnamefont{Yokosuk}},
  \bibinfo{author}{\bibfnamefont{A.}~\bibnamefont{Al-Wahish}},
  \bibinfo{author}{\bibfnamefont{S.}~\bibnamefont{Artyukhin}},
  \bibinfo{author}{\bibfnamefont{K.}~\bibnamefont{O’Neal}},
  \bibinfo{author}{\bibfnamefont{D.}~\bibnamefont{Mazumdar}},
  \bibinfo{author}{\bibfnamefont{P.}~\bibnamefont{Chen}},
  \bibinfo{author}{\bibfnamefont{J.}~\bibnamefont{Yang}},
  \bibinfo{author}{\bibfnamefont{Y.~S.} \bibnamefont{Oh}},
  \bibinfo{author}{\bibfnamefont{S.~A.} \bibnamefont{McGill}},
  \bibinfo{author}{\bibfnamefont{K.}~\bibnamefont{Haule}},
  \bibnamefont{et~al.}, \bibinfo{journal}{Phys. Rev. Lett.}
  \textbf{\bibinfo{volume}{117}}, \bibinfo{pages}{147402}
  (\bibinfo{year}{2016}).

\bibitem[{\citenamefont{Ivanov et~al.}(2013)\citenamefont{Ivanov, Mathieu,
  Nordblad, Tellgren, Ritter, Politova, Kaleva, Mosunov, Stefanovich, and
  Weil}}]{ivanov2013spin}
\bibinfo{author}{\bibfnamefont{S.~A.} \bibnamefont{Ivanov}},
  \bibinfo{author}{\bibfnamefont{R.}~\bibnamefont{Mathieu}},
  \bibinfo{author}{\bibfnamefont{P.}~\bibnamefont{Nordblad}},
  \bibinfo{author}{\bibfnamefont{R.}~\bibnamefont{Tellgren}},
  \bibinfo{author}{\bibfnamefont{C.}~\bibnamefont{Ritter}},
  \bibinfo{author}{\bibfnamefont{E.}~\bibnamefont{Politova}},
  \bibinfo{author}{\bibfnamefont{G.}~\bibnamefont{Kaleva}},
  \bibinfo{author}{\bibfnamefont{A.}~\bibnamefont{Mosunov}},
  \bibinfo{author}{\bibfnamefont{S.}~\bibnamefont{Stefanovich}},
  \bibnamefont{and} \bibinfo{author}{\bibfnamefont{M.}~\bibnamefont{Weil}},
  \bibinfo{journal}{Chem. Mater.} \textbf{\bibinfo{volume}{25}},
  \bibinfo{pages}{935} (\bibinfo{year}{2013}).

\bibitem[{\citenamefont{Ross et~al.}(2015)\citenamefont{Ross, Bordelon, Terho,
  and Neilson}}]{ross2015nanosized}
\bibinfo{author}{\bibfnamefont{K.~A.} \bibnamefont{Ross}},
  \bibinfo{author}{\bibfnamefont{M.}~\bibnamefont{Bordelon}},
  \bibinfo{author}{\bibfnamefont{G.}~\bibnamefont{Terho}}, \bibnamefont{and}
  \bibinfo{author}{\bibfnamefont{J.~R.} \bibnamefont{Neilson}},
  \bibinfo{journal}{Phys. Rev. B} \textbf{\bibinfo{volume}{92}},
  \bibinfo{pages}{134419} (\bibinfo{year}{2015}).

\bibitem[{\citenamefont{Bogdanov and
  Yablonskii}(1989)}]{bogdanov1989thermodynamically}
\bibinfo{author}{\bibfnamefont{A.}~\bibnamefont{Bogdanov}} \bibnamefont{and}
  \bibinfo{author}{\bibfnamefont{D.}~\bibnamefont{Yablonskii}},
  \bibinfo{journal}{Zh. Eksp. Teor. Fiz} \textbf{\bibinfo{volume}{95}},
  \bibinfo{pages}{182} (\bibinfo{year}{1989}).

\bibitem[{\citenamefont{Kimura et~al.}(2003)\citenamefont{Kimura, Goto,
  Shintani, Ishizaka, Arima, and Tokura}}]{2003magnetic}
\bibinfo{author}{\bibfnamefont{T.}~\bibnamefont{Kimura}},
  \bibinfo{author}{\bibfnamefont{T.}~\bibnamefont{Goto}},
  \bibinfo{author}{\bibfnamefont{H.}~\bibnamefont{Shintani}},
  \bibinfo{author}{\bibfnamefont{K.}~\bibnamefont{Ishizaka}},
  \bibinfo{author}{\bibfnamefont{T.}~\bibnamefont{Arima}}, \bibnamefont{and}
  \bibinfo{author}{\bibfnamefont{Y.}~\bibnamefont{Tokura}},
  \bibinfo{journal}{Nature (London)} \textbf{\bibinfo{volume}{426}},
  \bibinfo{pages}{55} (\bibinfo{year}{2003}).

\bibitem[{\citenamefont{Weil et~al.}(2014)\citenamefont{Weil, Mathieu,
  Nordblad, and Ivanov}}]{weil2014crystal}
\bibinfo{author}{\bibfnamefont{M.}~\bibnamefont{Weil}},
  \bibinfo{author}{\bibfnamefont{R.}~\bibnamefont{Mathieu}},
  \bibinfo{author}{\bibfnamefont{P.}~\bibnamefont{Nordblad}}, \bibnamefont{and}
  \bibinfo{author}{\bibfnamefont{S.}~\bibnamefont{Ivanov}},
  \bibinfo{journal}{Cryst. Res. Technol.} \textbf{\bibinfo{volume}{49}},
  \bibinfo{pages}{142} (\bibinfo{year}{2014}).

\bibitem[{\citenamefont{Prosnikov et~al.}(2019)\citenamefont{Prosnikov,
  Smirnov, Davydov, Araki, Arima, and Pisarev}}]{prosnikov2019lattice}
\bibinfo{author}{\bibfnamefont{M.}~\bibnamefont{Prosnikov}},
  \bibinfo{author}{\bibfnamefont{A.}~\bibnamefont{Smirnov}},
  \bibinfo{author}{\bibfnamefont{V.~Y.} \bibnamefont{Davydov}},
  \bibinfo{author}{\bibfnamefont{Y.}~\bibnamefont{Araki}},
  \bibinfo{author}{\bibfnamefont{T.}~\bibnamefont{Arima}}, \bibnamefont{and}
  \bibinfo{author}{\bibfnamefont{R.}~\bibnamefont{Pisarev}},
  \bibinfo{journal}{Phys. Rev. B} \textbf{\bibinfo{volume}{100}},
  \bibinfo{pages}{144417} (\bibinfo{year}{2019}).

\bibitem[{\citenamefont{Wang et~al.}(2015)\citenamefont{Wang, Huang, Yang, Oh,
  and Cheong}}]{wang2015interlocked}
\bibinfo{author}{\bibfnamefont{X.}~\bibnamefont{Wang}},
  \bibinfo{author}{\bibfnamefont{F.-T.} \bibnamefont{Huang}},
  \bibinfo{author}{\bibfnamefont{J.}~\bibnamefont{Yang}},
  \bibinfo{author}{\bibfnamefont{Y.~S.} \bibnamefont{Oh}}, \bibnamefont{and}
  \bibinfo{author}{\bibfnamefont{S.-W.} \bibnamefont{Cheong}},
  \bibinfo{journal}{APL Mater.} \textbf{\bibinfo{volume}{3}},
  \bibinfo{pages}{076105} (\bibinfo{year}{2015}).

\bibitem[{\citenamefont{Yamasaki et~al.}(2013)\citenamefont{Yamasaki, Sudayama,
  Okamoto, Nakao, Kubota, and Murakami}}]{yamasaki2013diffractometer}
\bibinfo{author}{\bibfnamefont{Y.}~\bibnamefont{Yamasaki}},
  \bibinfo{author}{\bibfnamefont{T.}~\bibnamefont{Sudayama}},
  \bibinfo{author}{\bibfnamefont{J.}~\bibnamefont{Okamoto}},
  \bibinfo{author}{\bibfnamefont{H.}~\bibnamefont{Nakao}},
  \bibinfo{author}{\bibfnamefont{M.}~\bibnamefont{Kubota}}, \bibnamefont{and}
  \bibinfo{author}{\bibfnamefont{Y.}~\bibnamefont{Murakami}},
  \bibinfo{journal}{J. Phys.: Conf. Ser.} \textbf{\bibinfo{volume}{425}},
  \bibinfo{pages}{132012} (\bibinfo{year}{2013}).

\bibitem[{\citenamefont{Yamasaki et~al.}(2015)\citenamefont{Yamasaki, Morikawa,
  Honda, Nakao, Murakami, Kanazawa, Kawasaki, Arima, and
  Tokura}}]{yamasaki2015dynamical}
\bibinfo{author}{\bibfnamefont{Y.}~\bibnamefont{Yamasaki}},
  \bibinfo{author}{\bibfnamefont{D.}~\bibnamefont{Morikawa}},
  \bibinfo{author}{\bibfnamefont{T.}~\bibnamefont{Honda}},
  \bibinfo{author}{\bibfnamefont{H.}~\bibnamefont{Nakao}},
  \bibinfo{author}{\bibfnamefont{Y.}~\bibnamefont{Murakami}},
  \bibinfo{author}{\bibfnamefont{N.}~\bibnamefont{Kanazawa}},
  \bibinfo{author}{\bibfnamefont{M.}~\bibnamefont{Kawasaki}},
  \bibinfo{author}{\bibfnamefont{T.}~\bibnamefont{Arima}}, \bibnamefont{and}
  \bibinfo{author}{\bibfnamefont{Y.}~\bibnamefont{Tokura}},
  \bibinfo{journal}{Phys. Rev. B} \textbf{\bibinfo{volume}{92}},
  \bibinfo{pages}{220421} (\bibinfo{year}{2015}).

\bibitem[{\citenamefont{Takata et~al.}(2015)\citenamefont{Takata, Suzuki,
  Shinohara, Oku, Tominaga, Ohishi, Iwase, Nakatani, Inamura, Ito
  et~al.}}]{takata2015design}
\bibinfo{author}{\bibfnamefont{S.-i.} \bibnamefont{Takata}},
  \bibinfo{author}{\bibfnamefont{J.-i.} \bibnamefont{Suzuki}},
  \bibinfo{author}{\bibfnamefont{T.}~\bibnamefont{Shinohara}},
  \bibinfo{author}{\bibfnamefont{T.}~\bibnamefont{Oku}},
  \bibinfo{author}{\bibfnamefont{T.}~\bibnamefont{Tominaga}},
  \bibinfo{author}{\bibfnamefont{K.}~\bibnamefont{Ohishi}},
  \bibinfo{author}{\bibfnamefont{H.}~\bibnamefont{Iwase}},
  \bibinfo{author}{\bibfnamefont{T.}~\bibnamefont{Nakatani}},
  \bibinfo{author}{\bibfnamefont{Y.}~\bibnamefont{Inamura}},
  \bibinfo{author}{\bibfnamefont{T.}~\bibnamefont{Ito}}, \bibnamefont{et~al.},
  \bibinfo{journal}{JPS Conf. Proc.} \textbf{\bibinfo{volume}{8}},
  \bibinfo{pages}{036020} (\bibinfo{year}{2015}).

\bibitem[{\citenamefont{Tokura et~al.}(2014)\citenamefont{Tokura, Seki, and
  Nagaosa}}]{tokura2014multiferroics}
\bibinfo{author}{\bibfnamefont{Y.}~\bibnamefont{Tokura}},
  \bibinfo{author}{\bibfnamefont{S.}~\bibnamefont{Seki}}, \bibnamefont{and}
  \bibinfo{author}{\bibfnamefont{N.}~\bibnamefont{Nagaosa}},
  \bibinfo{journal}{Rep. Prog. Phys.} \textbf{\bibinfo{volume}{77}},
  \bibinfo{pages}{076501} (\bibinfo{year}{2014}).

\bibitem[{\citenamefont{Pappas}(2012)}]{pappas2012new}
\bibinfo{author}{\bibfnamefont{C.}~\bibnamefont{Pappas}},
  \bibinfo{journal}{Physics} \textbf{\bibinfo{volume}{5}}, \bibinfo{pages}{28}
  (\bibinfo{year}{2012}).

\bibitem[{\citenamefont{Ederer and Spaldin}(2005)}]{ederer2005weak}
\bibinfo{author}{\bibfnamefont{C.}~\bibnamefont{Ederer}} \bibnamefont{and}
  \bibinfo{author}{\bibfnamefont{N.~A.} \bibnamefont{Spaldin}},
  \bibinfo{journal}{Phys. Rev. B} \textbf{\bibinfo{volume}{71}},
  \bibinfo{pages}{060401} (\bibinfo{year}{2005}).

\bibitem[{\citenamefont{Tokunaga
  et~al.}(2015{\natexlab{b}})\citenamefont{Tokunaga, Akaki, Ito, Miyahara,
  Miyake, Kuwahara, and Furukawa}}]{tokunaga2015magnetic}
\bibinfo{author}{\bibfnamefont{M.}~\bibnamefont{Tokunaga}},
  \bibinfo{author}{\bibfnamefont{M.}~\bibnamefont{Akaki}},
  \bibinfo{author}{\bibfnamefont{T.}~\bibnamefont{Ito}},
  \bibinfo{author}{\bibfnamefont{S.}~\bibnamefont{Miyahara}},
  \bibinfo{author}{\bibfnamefont{A.}~\bibnamefont{Miyake}},
  \bibinfo{author}{\bibfnamefont{H.}~\bibnamefont{Kuwahara}}, \bibnamefont{and}
  \bibinfo{author}{\bibfnamefont{N.}~\bibnamefont{Furukawa}},
  \bibinfo{journal}{Nat. Commun.} \textbf{\bibinfo{volume}{6}},
  \bibinfo{pages}{5878} (\bibinfo{year}{2015}{\natexlab{b}}).

\bibitem[{DMI()}]{DMI}
\bibinfo{note}{{T}he length of DM vector \mbox{\boldmath $D$}$_1$ is different
  from that of \mbox{\boldmath $D$}$_2$, because of the distinct local atom
  arrangement.}

\bibitem[{\citenamefont{Ji et~al.}(2018)\citenamefont{Ji, Solana, Lopez,
  Manuel, Ritter, Senyshyn, and Attfield}}]{ji2018lock}
\bibinfo{author}{\bibfnamefont{K.-L.} \bibnamefont{Ji}},
  \bibinfo{author}{\bibfnamefont{E.}~\bibnamefont{Solana}},
  \bibinfo{author}{\bibfnamefont{A.~M.~A.} \bibnamefont{Lopez}},
  \bibinfo{author}{\bibfnamefont{P.}~\bibnamefont{Manuel}},
  \bibinfo{author}{\bibfnamefont{C.}~\bibnamefont{Ritter}},
  \bibinfo{author}{\bibfnamefont{A.}~\bibnamefont{Senyshyn}}, \bibnamefont{and}
  \bibinfo{author}{\bibfnamefont{P.}~\bibnamefont{Attfield}},
  \bibinfo{journal}{Chem. Commun.} \textbf{\bibinfo{volume}{54}},
  \bibinfo{pages}{12523} (\bibinfo{year}{2018}).

\bibitem[{\citenamefont{Katsura et~al.}(2005)\citenamefont{Katsura, Nagaosa,
  and Balatsky}}]{katsura2005spin}
\bibinfo{author}{\bibfnamefont{H.}~\bibnamefont{Katsura}},
  \bibinfo{author}{\bibfnamefont{N.}~\bibnamefont{Nagaosa}}, \bibnamefont{and}
  \bibinfo{author}{\bibfnamefont{A.~V.} \bibnamefont{Balatsky}},
  \bibinfo{journal}{Phys. Rev. Lett.} \textbf{\bibinfo{volume}{95}},
  \bibinfo{pages}{057205} (\bibinfo{year}{2005}).

\bibitem[{\citenamefont{Arima}(2007)}]{arima2007ferroelectricity}
\bibinfo{author}{\bibfnamefont{T.}~\bibnamefont{Arima}}, \bibinfo{journal}{J.
  Phys. Soc. Jpn.} \textbf{\bibinfo{volume}{76}}, \bibinfo{pages}{073702}
  (\bibinfo{year}{2007}).

\end{thebibliography}
\end{document}